\documentclass[lettersize,journal]{IEEEtran}
\usepackage{amsmath,amsfonts}
\usepackage{algorithmic}
\usepackage{algorithm}
\usepackage{array}
\usepackage[caption=false,font=normalsize,labelfont=sf,textfont=sf]{subfig}
\usepackage{textcomp}
\usepackage{stfloats}
\usepackage{url}
\usepackage{verbatim}
\usepackage{graphicx}
\usepackage{cite}
\usepackage{multirow}
\usepackage{xcolor}
\begin{document}

\title{A Tree-guided CNN for image super-resolution}

\author{Chunwei Tian,~\IEEEmembership{Member,~IEEE,} Mingjian Song, Xiaopeng Fan,~\IEEEmembership{Senior Member,~IEEE,} Xiangtao Zheng, Bob Zhang,~\IEEEmembership{Senior Member,~IEEE,} and David Zhang,~\IEEEmembership{Life Fellow,~IEEE}
\thanks{Manuscript received XX, XXXX; revised XX XX, XXXX. This work was supported in part by the National Natural Science Foundation of China (NSFC) under Grant 62201468, U22B2035 and 62441202, in part by the University of Macau MYRG-GRG2024-00205-FST-UMDF, and in part by the National Key R\&D Program of China under Grant 2021YFF0900500. \textit{(Corresponding author: Xiaopeng Fan and Bob Zhang.)}}
\thanks{Chunwei Tian is with School of Computer Science and Technology, Harbin Institute of Technology, Harbin 150001, China (e-mail: chunweitian@hit.edu.cn).}
\thanks{Mingjian Song is with School of Software, Northwestern Polytechnical University, Xi’an 710129, China (e-mail: songmingjian@mail.nwpu.edu.cn).}
\thanks{Xiaopeng Fan is with School of Computer Science and Technology, Harbin Institute of Technology, Harbin 150001, China (e-mail: fxp@hit.edu.cn).}
\thanks{Xiangtao Zheng is with College of Physics and Information Engineering, Fuzhou University, Fuzhou 350108, China (e-mail: xiangtaoz@gmail.com).}
\thanks{Bob Zhang is with Pattern Analysis and Machine Intelligence Group, Department of Computer and Information Science, University of Macau, Macau 999078, China (e-mail: bobzhang@um.edu.mo).}
\thanks{David Zhang is with the School of Data Science, The Chinese University of Hong Kong (Shenzhen), Shenzhen, 518172, Guangdong, China. And he is also with the Shenzhen Institute of Artiﬁcial Intelligence and Robotics for Society, Shenzhen, China (Email: davidzhang@cuhk.edu.cn)}}

\markboth{Journal of IEEE Transactions on Consumer Electronics,~Vol.~XX, No.~X, XX~XXXX}%
{Tian \MakeLowercase{\textit{et al.}}: A Tree-guided CNN for image super-resolution}

\IEEEpubid{0000--0000/00\$00.00~\copyright~2021 IEEE}

\maketitle

\begin{abstract}
Deep convolutional neural networks can extract more accurate structural information via deep architectures to obtain good performance in image super-resolution. However, it is not easy to find effect of important layers in a single network architecture to decrease performance of super-resolution. In this paper, we design a tree-guided CNN for image super-resolution (TSRNet). It uses a tree architecture to guide a deep network to enhance effect of key nodes to amplify the relation of hierarchical information for improving the ability of recovering images. To prevent insufficiency of the obtained structural information, cosine transform techniques in the TSRNet are used to extract cross-domain information to improve the performance of image super-resolution. Adaptive Nesterov momentum optimizer (Adan) is applied to optimize parameters to boost effectiveness of training a super-resolution model. Extended experiments can verify superiority of the proposed TSRNet for restoring high-quality images. Its code can be obtained at \url{https://github.com/hellloxiaotian/TSRNet.} 
\end{abstract}


\section{Introduction}
\IEEEPARstart{I}{mage} super-resolution (SR) utilized a degradation model to achieve a mapping from unclear images to high-quality images \cite{yang2019deep}. Due to different damaged degree, image super-resolution problem has many solutions, which is also an ill-posed problem. To overcome drawbacks above, some traditional machine learning methods based pixels, i.e., nearest neighbor interpolation \cite{rukundo2012nearest}, bilinear interpolation \cite{li2001new} and bicubic interpolation \cite{keys1981cubic} were proposed for image super-resolution. Although these methods are easy to be implemented, their super-resolution performance is suboptimal \cite{park2003super}. To obtain more detailed information, priori knowledge, i.e., sparse representation \cite{liao2023minimax} is used to facilitate more high-frequency structural information for image super-resolution \cite{yang2010image}. For instance, Yang et al. utilized sparse representation and dictionary learning to accurately restore high-frequency information to improve quality of predicted images \cite{yang2008image}. Alternatively, Rosseau et al. used non-local similarity constraints to mine more structure and texture information to improve effect of super-resolution \cite{rousseau2010non}.  Although these methods can effectively deal with image super-resolution, they need to refer to complex sophisticated optimization techniques, which increased computational costs for finding optimal parameters. Besides, the mentioned methods depended on manually setting parameters to obtain optimal solution for image super-resolution. 

To overcome problem above, deep learning techniques can extract accurate structural information via deep network architectures for recovering high-quality images \cite{dong2014learning}, which have become popular tools in image super-resolution \cite{tian2020coarse}. Specifically, convolutional neural networks (CNNs) with end-to-end architectures can make users design flexibly networks to meet their demands, which has become a mainstream method for image super-resolution \cite{dong2015image}. Dong et al. designed a 3-layer network architecture of SRCNN via pixel mapping ways to recover high-quality images from low-resolution images \cite{dong2014learning}. Although SRCNN has obtained more exciting effects of super-resolution compared with classical methods, it has poor extension ability. To address this problem, stacked convolutional layers can extract more useful structural information in image super-resolution \cite{hu2019runet}. Kim et al. cascaded convolutional layers and ReLU via residual learning operations to expand network depth to enlarge perception field for image super-resolution \cite{kim2016accurate}. Although previous deep networks have achieved excellent results, they still face challenges of referring to more parameters. To solve this mentioned problem, Kim at el. combined recursive layer and skip connection into a CNN to remove unnecessary parameters for image super-resolution \cite{kim2016deeply}.  Although these methods can performance well for image super-resolution, they used bicubic interpolation operations to amplify given low-resolution images as inputs of CNNs, which may cause higher computational costs.  

\IEEEpubidadjcol
To address this problem, an up-sampling operation in a deep layer is presented to enlarge obtained low-frequency information to reduce computational costs in image SR \cite{dong2016accelerating}. For instance, Dong et al. customized a deconvolutional layer at end of a CNN rather than original bicubic interpolation at start of a CNN to obtain high-frequency information for SR, which can make the designed CNN keep low computational costs in SR \cite{dong2016accelerating}. To improve relations of key salient information, asymmetric convolutions are used to improve effects of local information in horizontal and vertical directions to overcome long-term dependency problem and improve SR effect \cite{tian2021asymmetric}. To overcome the expansion of parameters for modeling long-term dependency, a proposed local concatenation learning strategy utilized a lightweight depth-wise separable convolution layer to remove redundant information from feature maps after concatenation between low-level and high-level layers to ensure efficient feature refinement and reduce complexity \cite{wanga2023hybrid}. Although these methods have obtained good SR results, they do not make full use of key layers, which can make an obtained SR performance be limited. 

In this paper, we used a tree idea to guide a CNN in image super-resolution (TSRNet). TSRNet utilizes a tree network architecture to enhance the relation of hierarchical information to improve its ability of restoring high-quality images, according to a binary tree idea. To prevent insufficiency of the obtained structural information, cosine transform technique is used to facilitate local salient information for image SR. To further suppress gradient explosion, adaptive Nesterov momentum algorithm is employed to optimize parameters to improve SR performance. The contributions can be summarized as follows: 

(1) A tree network is designed to enhance the relation of hierarchical information to improve SR performance, according to a binary tree.  

(2) A cosine transform technique is used to extract local salient information to facilitate more robust structural information in image super-resolution.  

(3) Adaptive Nesterov momentum algorithm is applied to optimize parameters to further suppress gradient explosion to improve performance of image SR. 

The following sections of our paper are shown: Section II summaries the related work of enhanced CNNs with designing new architectures for image super-resolution and deep networks with optimization methods for image super-resolution. Section III introduces the proposed method. Section IV illustrates experimental datasets, settings, analysis and results in image super-resolution. Section V concludes this paper. 

\section{Related work}
\subsection{Enhanced CNNs for image super-resolution}
CNNs can utilize deep architectures in feedforward and feedback ways to obtain powerful learning capability in image super-resolution. The methods based CNNs can be roughly divided into two categories: those that first interpolate the low-resolution image input before applying a learning-based super-resolution model, and those that directly learn the mapping from the original low-resolution image to the high-resolution output.

For the first method, scholars utilized interpolation operation, i.e., bicubic interpolation operation to enlarge given unclear images as same size with given reference high-quality images to input a CNN to learn a mapping relation in image SR \cite{fan2018compressed}. For instance, stacked some smaller convolutional layers can enlarge perception field to improve image SR \cite{kim2016accurate}. To address long-term dependency problem, Tai et al. introduced a memory block via a recursive unit and a gate unit to keep memory to obtain adaptive adjusting ability of shallow layers for image SR \cite{tai2017memnet}.  Alternatively, multiple residual units were used into a deep network by sharing weights to restore clearer images \cite{tai2017image}. Although these methods have performed well in image SR, they may have bigger computational costs by same network input and output sizes. 

For the second method, scholars utilized an up-sampling operation, i.e., transposed convolutional layer, sub-pixel layer and meta upscale module in a deep layer to enlarge low-frequency information to predict high-quality images \cite{wang2020deep}. For instance, Tong et al. introduced dense skip-connections in a deep network to jointly extract low-frequency information and used a transposed convolutional layer to capture more obtained low-frequency information to obtain high-resolution images \cite{tong2017image}. Alternatively, an information multi distillation network gradually integrated obtained low-frequency information to obtain more robust features for image SR \cite{hui2019lightweight}. To further enhance the expressive power of deep convolutional neural networks, numerous networks are designed for image SR. Ledig et al. used generative adversarial networks (GANs) with a perceptual loss to obtain visually more realistic image by a generator and discriminator \cite{ledig2017photo}. Gao et al. used a structure attention and a Transformer block to extract global and local receptive field to improve performance of image SR \cite{gao2023ctcnet}. Although they are effective in image SR, how to protect efficient SR performance via fully important layers has a challenge in image SR. In this paper, we use binary tree idea to find key nodes to guide a CNN for implementing an efficient super-resolution method.

\subsection{Deep Networks with optimization methods for image super-resolution}
Deep CNNs have powerful modeling and feature extraction capabilities in image SR. However, during the training process, deeper networks may suffer from gradient explosion or vanishing, slow convergence speed and local optimum to decrease performance of image SR. To solve these problems, researchers proposed combinations of optimization techniques and deep networks for image SR. These methods can commonly be divided into two categories: optimizing loss function and improving optimizer. 

Optimizing loss function can obtain more detailed information for image SR. For instance, Bulat et al. used a pretrained face alignment network to obtain prior knowledge of facial alignment, and jointly trained super-resolution network to improve performance of facial image super-resolution \cite{bulat2018super}. Besides, improving L1 and L2 loss is effective for image super-resolution \cite{wang2020deep}.  Lai et al. employed Charbonnier based L1 loss to penalize every level in their Laplacian pyramid network to improve convergence in image SR \cite{lai2017deep}. To improve visual quality of obtained super-resolution images, Johnson et al. developed a perceptual loss to train a network to improve quality of predicted images \cite{johnson2016perceptual}. Although these methods can extract complementary information with CNNs, they may consume more time. 

Deeper network often may suffer from gradient explosion, vanishing for image SR. 
To overcome this problem, various optimizers, i.e., gradient based optimizers, gradient accelerated optimizers and adaptive learning rate optimizers are developed. Gradient based optimizers and accelerated optimizers, i.e. Stochastic Gradient Descent (SGD) \cite{robbins1951stochastic}, are usually suitable to shallow networks and have faster optimization speeds. Early SRCNN utilizes SGD to minimize the loss to overcome long-term dependency problem \cite{dong2014learning}. However, this optimizer may limit the speed of training model for image SR \cite{dong2016accelerating}. To deal with this problem, a very deep convolutional network for image super-resolution utilizes a gradient accelerated optimizer based on back propagation and adjustable gradient clipping to improve training speed for image SR \cite{kim2016accurate}. Besides, as the modules used in the network become diverse and the branches of the overall network structure become complex, simple gradient descent optimization cannot make loss function converge well. To better optimize different components of the whole network, Lim et al. used Adaptive Moment Estimation (Adam) \cite{kingma2014adam} in their enhanced network to adjust learning rate of different parameters to find optimal parameters for image SR \cite{lim2017enhanced}.  SRGAN used an Adam to eliminate gradient vanishing and explosion for image SR \cite{ledig2017photo}.  Loshchilov et al. separated weight decay from optimization steps related to loss function to restore original regularized form weight decay to improve Adam’s generalization performance in image SR \cite{loshchilov2017decoupled}. Xie et al. proposed Adan to avoid local optimum of Adam for image SR \cite{xie2024adan}. To design most adaptive optimization strategy, we apply Adan for a designed binary tree network to further improve performance of image SR. 
\section{The proposed method}
In this section, we first introduce the network architecture of our TSRNet as shown in Fig \ref{fig1}. Then, we show key techniques of our TSRNet including a loss function, a cosine transform mechanism and optimization strategy in image SR.

\subsection{Network architecture}

\begin{figure*}[!t]
\centering
\includegraphics[width=\textwidth]{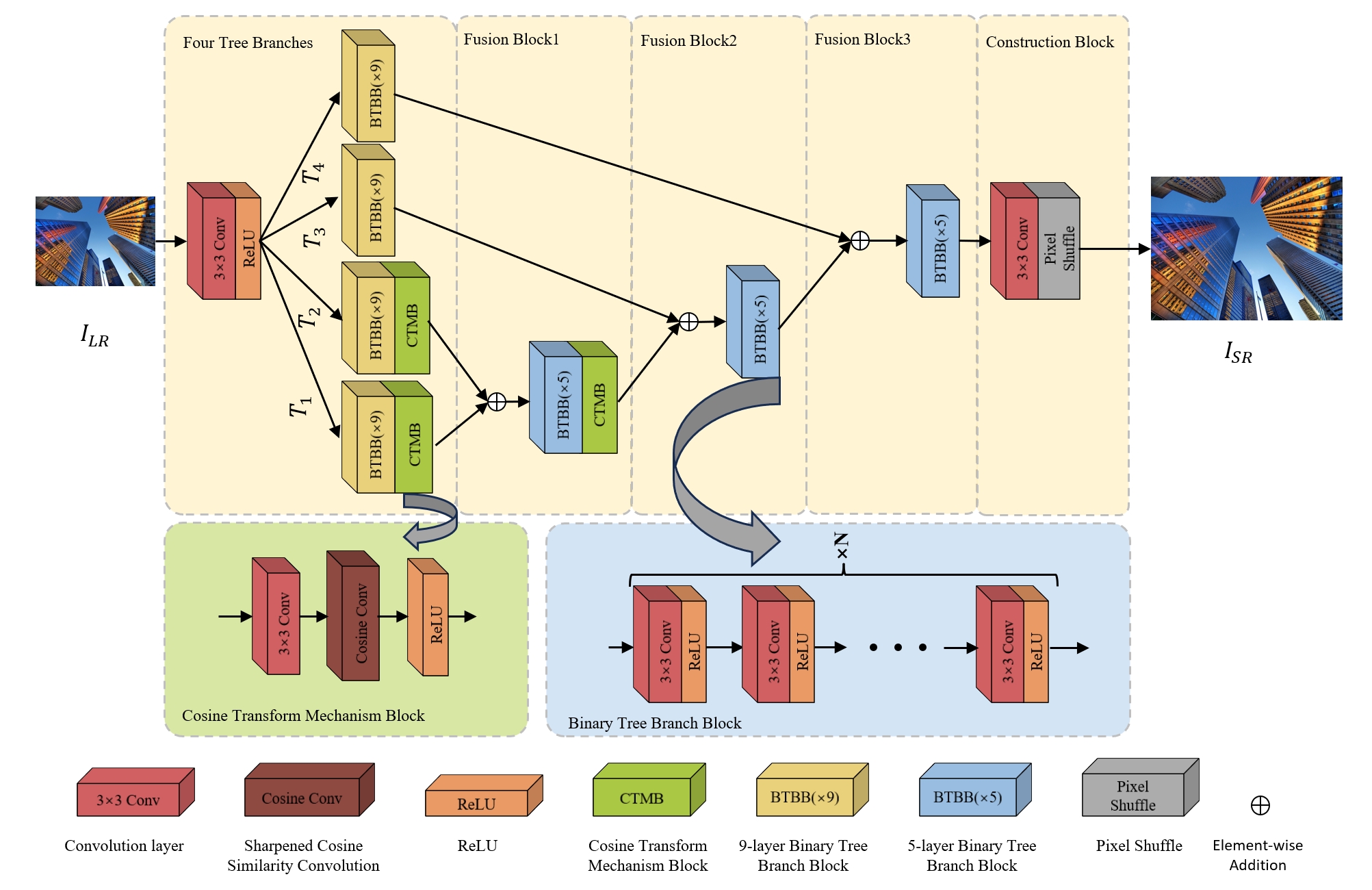}
\caption{Network architecture of the proposed TSRNet.}
\label{fig1}
\end{figure*}

The proposed TSRNet for image super-resolution comprises a tree architecture, cosine transform mechanism and optimization strategy for training process as shown in Fig \ref{fig1}. The entire process consists of four tree branches, three fusion blocks and one reconstruction block. Specifically, four tree branches containing a binary tree branch block (BTBB) and cosine transform mechanism blocks (CTMBs) use fusion blocks to enhance relation between different tree branches to extract complementary structural information for SR. Three fusion blocks sequentially fuse the output feature mappings of four branches by addition CTMB and BTBB to refine obtained structural information of different key node layers. That is, outputs of the first and second trees are gathered by fusion block1, then, obtained output and output of the third tree are gathered by fusion block2. Finally, obtained output and output of the third tree are gathered by fusion block3. To obtain high-quality images, a reconstruction block is used to construct high-resolution images. Specifically, the number of BTBB of different layer is different, which can be shown below. Also, CTMB can be introduced in Section III.C. The above process can be represented via the following formula \eqref{eq1}.
\begin{equation}
\begin{aligned}
I_{SR} &= TSRNet(I_{LR}) \\
       &= CB(f_3(f_2(f_1(T_1(I_{LR}), T_2(I_{LR})), T_3(I_{LR})), T_4(I_{LR})))
\label{eq1}
\end{aligned}
\end{equation}
where $I_{LR}$ denotes a low-resolution input image of TSRNet. $TSRNet$ represents a function of TSRNet. $I_{SR}$ is a super-resolution output image of TSRNet. $T_1$, $T_2$, $T_3$ and $T_4$ stand for functions of the first, second, third, fourth tree branches, respectively. $f_1$, $f_2$ and $f_3$ denote functions of the first, second and third fusion mechanisms, respectively. $CB$ is a construction block. $T_1$ and $T_2$ include a combination of a convolutional layer with $3\times3$, a ReLU, 9BTBB and a CTMB. $T_3$ and $T_4$ are composed of stacking a convolutional layer with $3\times3$, a ReLU and 9BTBB. 9BTBB contains nine stacked layers in the BTBB, which contains a convolutional layer and a ReLU. 

Functions of $T_1$, $T_2$, $T_3$ and $T_4$ can be shown as follows. 
\begin{equation}
\begin{aligned}
T_1(I_{LR}) = T_2(I_{LR}) &= CTMB(9BTBB(R(Conv(I_{LR}))))\\
                          &= CTMB(9R(Conv(R(Conv(I_{LR})))))
\label{eq2}
\end{aligned}
\end{equation}
\begin{equation}
\begin{aligned}
T_3(I_{LR}) = T_4(I_{LR}) &= 9BTBB(R(Conv(I_{LR})))\\
                          &= 9R(Conv(R(Conv(I_{LR}))))
\label{eq3}
\end{aligned}
\end{equation}
where $Conv$ stands for a function of a convolutional layer. $R$ is a function of ReLU. $BTBB$ stands for a function of a BTBB. $9BTBB$ is nine stacked layer in the BTBB. $9R(Conv)$ is nine stacked combinations of a convolutional layer and ReLU. $CTMB$ stands for a function of CTMB, which is shown in Section III.C. 

To gather information from different branches, three fusion mechanisms are used to merge obtained structural information of four branches for image super-resolution. Specifically, the first fusion mechanism as well as $f_1$ is composed of a residual operation, five BTBBs and a CTMB, where a residual operation is acted between $T_1$ and $T_2$. The second fusion mechanism of $f_2$ and third fusion mechanism of $f_3$ contains five BTBBs. Also, $f_2$ is used to act between $f_1$ and $T_3$. $f_3$ is used to act between $f_2$ and $T_4$. That process can be figuratively expressed via Eqs.\eqref{eq4}-\eqref{eq6}.
\begin{align}
f_1 &= CTMB(5BTBB(T_1 + T_2))\label{eq4}\\
f_2 &= 5BTBB(f_1 + T_3)\label{eq5}\\
f_3 &= 5BTBB(f_2 + T_4)\label{eq6}
\end{align}
where $5BTBB$ stands for five combinations of convolutional layer and ReLU in the BTBB and  
$+$ denotes a residual learning operation.  

To obtain high-quality super-resolution images, a construction block of $CB$ is designed. It is composed of a convolutional layer and a pixel shuffle. The process can be shown as follows. 
\begin{equation}
\begin{aligned}
I_{SR} = CB = PS(Conv)\label{eq7}
\end{aligned}
\end{equation}
where $PR$ denotes a function of a pixel-shuffle function, which is utilized to amplify obtained low-frequency structural information to high-frequency structural information. Sizes of mentioned convolutional layers are $3\times3$.
\subsection{Loss function}
Due to the wide application of MSE for image super-resolution \cite{tian2022image}, it is used as a loss function to optimize parameters in this paper. Let $\left\{I_{LR}^{i}, I_{HR}^{i}\right\}_{i=1}^{n}$ denote  paired low- and high-resolution images from a trainset in Section IV.A. $I_{LR}^{i}$ and $I_{HR}^{i}$ are the $i-th$ low- and high-resolution images, respectively. The process above can be conducted as Eq. (\ref{eq8}).
\begin{equation}
\begin{aligned}
l(\theta) = \frac{1}{2n} \sum_{i=1}^{n} \left\| TSRNet(I_{LR}^{i}) - I_{HR}^{i} \right\|^2
\label{eq8}
\end{aligned}
\end{equation}
where $L$ represents a loss function of MSE and $\theta$ is parameters. 
\subsection{Cosine transform mechanism block}
CTMB is used to extract local salient information for image super-resolution, which contains a convolutional layer, a cosine convolutional layer \cite{pisoni2022sharpenedcossim} and a ReLU. The convolutional layer is used to extract linear structural information. Also, a cosine convolutional layer is used to compute similarity of different direction to obtain local salient information. A ReLU is utilized to transform non-linear information. The process can be formulated as Eq. (\ref{eq9}).
\begin{equation}
\begin{aligned}
CTMB = R(CS(Conv))\label{eq9}
\end{aligned}
\end{equation}
where $CS$ denotes a cosine convolution. 
Differing from regular convolutions, cosine convolution uses Sharpen Cosine Similarity (SCS) to replace the dot product operation to extract local salient information \cite{pisoni2022sharpenedcossim}. That at, firstly, numerator uses dot product to calculate similarity and the denominator divides the normalized term by the feature input and convolution kernel weights to reduce scale size sensitivity. This allows cosine convolution to focus on more directional features rather than absolute size. Additionally, SCS further introduces learnable parameters and to enable cosine convolution to automatically adjust the sensitivity of similarity. It can be represented by the following formula:
\begin{equation}
\begin{aligned}
SCS(f, k) = sign(f \cdot k) \left( \frac{f \cdot k}{\left( \| f \| + \epsilon \right) \left( \| k \| + \epsilon \right)} \right)^p\label{eq10}
\end{aligned}
\end{equation}
where $f$ is the input feature. $k$ is the weight of convolutional kernel. $\| f \|$ and $\| k \|$ stand for L2 norm of input feature map and the weight of convolution kernel, respectively. $\epsilon$ denotes a small positive learnable parameter. $p$ denotes a learnable parameter used to control the sensitivity of similarity. $(\cdot)$ represents dot product. $sign(\cdot)$ represents the symbol of dot product. 
\subsection{Optimization strategy for training process}
To further suppress gradient vanishing of a deeper network for image super-resolution, Adan \cite{xie2024adan} is used to further optimize parameters. That is, Adan utilizes Nesterov momentum to calculate the first and second moments of gradients to improve the convergence speed and obtain better solutions of our TSRNet. It can be represented via the following formulas, according to Ref.\cite{xie2024adan}. Eq. (\ref{eq11}) is used to update momentum, Eq. (\ref{eq12}) is used to update gradient different momentum, Eq. (\ref{eq13}) is used to update momentum of squared gradient, Eq. (\ref{eq14}) is used to calculate learning rate, Eq. (\ref{eq15}) is used to update parameters.
\begin{align}
p_j &= (1 - \alpha_1) p_{j-1} + \alpha_1 g_j\label{eq11}\\
q_j &= (1 - \alpha_2) q_{j-1} + \alpha_2 (g_j - g_{j-1})\label{eq12}\\
s_j &= (1 - \alpha_3) s_{j-1} + \alpha_3 \left( g_j + (1 - \alpha_2) (g_j - g_{j-1}) \right)^2\label{eq13}\\
r_j &= \frac{r}{\sqrt{\bar{s}_j} + \epsilon}\label{eq14}\\
w_{j+1} &= \frac{1}{1 + \lambda_j r} \left( w_j - r_j \circ \left( p_j + (1 - \alpha_2) q_j \right) \right)\label{eq15}
\end{align}
where $p_j$, $q_j$, $s_j$, $r_j$ and $w_j$ denote momentum, gradient difference momentum, momentum of squared gradient, recurrent learning rate and parameters, respectively. $\alpha_1$, $\alpha_2$ and $\alpha_3 \in [0, 1]$ denote momentum coefficient, which is a hyperparameter set manually. $r$ stands for learning rate. $\epsilon$ is a small positive number used to avoid zero denominators. $\lambda_j$ is weight decay as well as a positive number. 
\section{Experimental analysis and results}
\subsection{Datasets}
To comprehensively evaluate our model for image SR, public datasets are selected to train and test our TSRNet. In terms of training dataset, due to high quality, large quantity and diversity images, public DIV2K \cite{agustsson2017ntire} is used to ensure reproducibility and robustness of training image super-resolution model. DIV2K dataset is composed of 1000 high-quality color images from various scenes, i.e., people, cities, handmade objects and natural sceneries \cite{agustsson2017ntire}. 800 images are chosen as a training dataset. Each image includes three scale factors of $\times 2$, $\times 3$ and $\times 4$. To increase the number of training dataset, each single image is cropped as size of $128 \times 128$. 
In terms of test datasets, we select four popular datasets, including Set5 \cite{bevilacqua2012low}, Set14 \cite{zeyde2012single}, BSD100 (B100) \cite{martin2001database} and Urban100 (U100) \cite{huang2015single} to equitably compare and evaluate model performance in image super-resolution. Specifically, Set5 and Set14 have 5 and 14 natural images, respectively. B100 collects 100 natural images. U100 contains 100 images, where the urban structure of images can provide an effective assessment of the models’ capability to capture high-frequency by the presence of numerous edges and textures in urban scenes \cite{tian2020coarse}. All images of those datasets are resized in scales of $\times 2$, $\times 3$ and $\times 4$. Y channel of predicted images is used to test performance of our TSRNet for image SR. 
\subsection{Parameters settings}
To train our model, the epochs and batch size are 1,200 and 64, respectively. Also, we utilize the multi-step decay strategy to make training process more stable. Specifically, original learning rate is 4e-4 and decreases to half after 200 epochs. Also, we exploit Adan instead of Adam to optimize the loss function. Specially, the parameters of our Adan are $\beta_1$ of 0.98, $\beta_2$ of 0.92, $\beta_3$ of 0.99 and $\epsilon$ of $1e-8$. The parameter settings of all comparative experiments are consistent with the above. In addition, the parameters of Adam in ablation experiment are $\beta_1$ of 0.9, $\beta_2$ of 0.99 and $\epsilon$ of $1e-8$. The specific settings of ablation experiment can be seen in subsection C.

In terms of hardware and environmental settings, we carried out all experiments on PC with the operation system of Ubuntu 20.04. Specifically, we utilize 24G memory GPU of Nvidia GeForce RTX 3090 and CPU of AMD EPYC 7502P 32-Core Processor. Additionally, we use Pytorch1.13.1 and Python3.8 to train and test our model in image super-resolution. Also, CUDA11.7 and cuDNN8.5.0 can improve training speed of our TSRNet on GPU.
\begin{table}
\begin{center}
\caption{Mean PSNR results for different methods with $\times 4$ on U100.}
\label{tab1}
\begin{tabular}{| l | c |}
\hline
Methodologies & PSNR(dB)\\
\hline
TSRNet without CTMB, Adan, T1 , T2 and T3 & 25.19 \\
\hline
TSRNet without CTMB, Adan, T1 and T2 & 25.61\\
\hline
TSRNet without CTMB, Adan and T1 & 25.80\\ 
\hline
TSRNet without CTMB and Adan & 25.89\\
\hline 
TSRNet without CTMB & 25.97\\
\hline 
TSRNet (Ours) & 26.00\\
\hline
\end{tabular}
\end{center}
\end{table}
\subsection{Ablation study}
This paper can be optimized by designing efficient network and using advanced
optimizer for image super-resolution. In terms of designing efficient network, we design a tree architecture to mine representative structural information in image super-resolution. Also, using cosine transform technique is used to extract local salient information. Its validity and rationality can be shown as follows.

Deep networks can mine accurate structural information via each layer to recover detailed information in image SR. However, non-key layers may cover effects of key layers to decrease super-resolution effects. Thus, improving key layers is very important for image SR. Common SR methods can depend on increasing network depth \cite{kim2016accurate} or enlarging width \cite{wang2024multi} of deep networks to improve perception field to improve SR performance. However, deeper networks may increase computational costs and they are difficult to be trained. Also, wider networks may cause overfitting. To address these problems, residual dense network methods via fully using hierarchical information are presented for image SR \cite{zhang2018residual}. Although it can obtain good effect for image SR, it may face higher computational costs. To mine effects of key layers for image SR, we design a tree network architecture, according to binary tree idea \cite{yi2018separable} and deep learning theory \cite{tian2022image}. 

According to binary tree idea, we design four trees, i.e., T1, T2, T3 and T4 to enhance relation of key nodes for image SR, which can be illustrated in Section III. Three fusion mechanisms can enhance relations of four trees. That is, fusion block1 can gather T1 and T2. Fusion block2 can gather outputs of fusion block1 and T3. Fusion block3 can gather outputs of fusion block2 and T4. In terms of deep learning theory, increasing network depth can capture more content information for image applications \cite{kim2016accurate}. Thus, each tree stacked nine convolutional layers and ReLU to extract more accurate structural information for image SR. To prevent redundancy information from gathered different trees, stacked convolutional network can address this question \cite{tian2022heterogeneous}. Inspired by that, five stacked convolutional layers and ReLU are used to refine obtained structural information in three fusion mechanisms. Their effectiveness can be proved as follows. 

TSRNet without CTMB, Adan, T1 and T2 can obtain better effect than that of TSRNet without CTMB, Adan, T1, T2 and T3 for image SR in TABLE \ref{tab1}, which shows effectiveness of fusion of T3 and T4. TSRNet without CTMB, Adan and T1 can obtain better effect than that of TSRNet without CTMB, Adan, T1 and T2 for image SR in TABLE \ref{tab1}, which shows effectiveness of fusion of T2, T3 and T4. TSRNet without CTMB and Adan can obtain better effect than that of TSRNet without CTMB, Adan and T1 for image SR in TABLE \ref{tab1}, which shows effectiveness of fusion of T1, T2, T3 and T4. To prevent loss of local salient information, cosine convolutional blocks are used into two trees, which has two reasons. Firstly, cosine convolutional blocks can extract directional salient information to improve performance of image application \cite{pisoni2022sharpenedcossim}.  Secondly, heterogeneous architectures can facilitate more robust information for image SR \cite{tian2022heterogeneous}.  Thus, cosine convolutional blocks are used into a T1, T2 and fusion block1 to achieve a heterogeneous network to obtain more representative detailed information for image SR. Its effectiveness can be verified by TSRNet and TSRNet without CTMB in TABLE \ref{tab1}. 

To further improve performance, we use Adan \cite{xie2024adan} to further suppress gradient explosion to optimize parameters to improve performance of image SR. The results in TABLE \ref{tab1} indicate that the inclusion of Adan improves the performance of TSRNet. Specifically, TSRNet without CTMB achieves a PSNR improvement of 0.08 dB compared to  TSRNet without CTMB and Adan.

\begin{table}
\begin{center}
\caption{PSNR and SSIM results for different methods with different scales on Set5.}
\label{tab2}
\begin{tabular}{>{\centering\arraybackslash}p{0.8cm} >{\centering\arraybackslash}p{1.8cm} >{\centering\arraybackslash}p{1.24cm} >{\centering\arraybackslash}p{1.24cm} >{\centering\arraybackslash}p{1.24cm}}
\hline
\multirow{3}{*}{Dataset} & \multirow{3}{*}{Methods} 
& $\times 2 $ & $\times 3 $ & $\times 4 $ \\\cline{3-5}
& & PSNR(dB)& PSNR(dB) & PSNR(dB) \\
& & /SSIM & /SSIM & /SSIM \\\hline
\multirow{33}{*}{Set5} 
& Bicubic & 33.66/0.9299 & 30.39/0.8682 & 28.42/0.8104 \\
& SRCNN\cite{dong2014learning} & 36.66/0.9542 & 32.75/0.9090 & 30.48/0.8628 \\
& VDSR\cite{kim2016accurate}& 37.53/0.9587 & 33.66/0.9213 & 31.35/0.8838 \\
& DRRN\cite{tai2017image}& 37.74/0.9591 &34.03/0.9244 &31.68/0.8888\\
&FSRCNN\cite{dong2016accelerating}&37.00/0.9558 &33.16/0.9140 &30.71/0.8657\\ 
&CARN-M\cite{ahn2018fast}&37.53/0.9583 &33.99/0.9236 &31.92/0.8903\\
&IDN\cite{hui2018fast}&37.83/0.9600 &34.11/\textcolor{blue}{0.9253} &31.82/0.8903\\ 
&A+\cite{timofte2015a+}&36.54/0.9544 &32.58/0.9088 &30.28/0.8603\\
&JOR\cite{dai2015jointly}&36.58/0.9543 &32.55/0.9067 &30.19/0.8563 \\
&RFL\cite{schulter2015fast}&36.54/0.9537 &32.43/0.9057 &30.14/0.8548\\
&SelfEx\cite{huang2015single}&36.49/0.9537 &32.58/0.9093 &30.31/0.8619 \\
&CSCN\cite{wang2015deep}&36.93/0.9552 &33.10/0.9144 &30.86/0.8732\\
&RED\cite{mao2016image}&37.56/0.9595 &33.70/0.9222 &31.33/0.8847 \\
&MAP\cite{zhang2012single}&37.58/0.9590 &33.75/0.9222 &31.40/0.8845 \\
&TNRD\cite{chen2016trainable}&36.86/0.9556 &33.18/0.9152 &30.85/0.8732\\
&FDSR\cite{lu2018fast}&37.40/0.9513 &33.68/0.9096 &31.28/0.8658 \\
&RCN\cite{shi2017structure}&37.17/0.9583 &33.45/0.9175 &31.11/0.8736\\
&DRCN\cite{kim2016deeply}&37.63/0.9588 &33.82/0.9226 &31.53/0.8854 \\
&CNF\cite{ren2017image}&37.66/0.9590 &33.74/0.9226 &31.55/0.8856 \\
&LapSRN\cite{lai2017deep}&37.52/0.9590&-&31.54/0.8850\\
&WaveResNet\cite{bae2017beyond}&37.57/0.9586&33.86/0.9228&31.52/0.8864\\
&CPCA\cite{xu2018self}&34.99/0.9469&31.09/0.8975&28.67/0.8434\\
&NDRCN\cite{cao2019new}&37.73/0.9596&33.90/0.9235&31.50/0.8859\\
&MemNet\cite{tai2017memnet}&37.78/0.9597 &34.09/0.9248 &31.74/0.8893\\
&LESRCNN\cite{tian2020lightweight}&37.65/0.9586 &33.93/0.9231 &31.88/0.8903\\
&LESRCNN-S\cite{tian2020lightweight}&37.57/0.9582&34.05/0.9238&31.88/0.8907\\\
&DSRCNN\cite{song2022dual}&37.73/0.9588&34.17/0.9247&31.89/0.8909\\
&DAN\cite{huang2020unfolding}&37.34/0.9526&34.04/0.9199&31.89/0.8864\\
&DCLS\cite{luo2022deep}&37.63/0.9554&34.21/0.9218&\textcolor{red}{32.12}/0.8890\\
&PGAN\cite{shi2023structure}&-&-&31.03/0.8798\\
&AFAN-S\cite{wang2023image}&\textcolor{blue}{37.84}/\textcolor{blue}{0.9601}&\textcolor{red}{34.22}/\textcolor{blue}{0.9253}&\textcolor{blue}{32.04}/\textcolor{blue}{0.8914}\\
&ACDMSR\cite{niu2024acdmsr}&36.46/0.9431&33.00/0.9059&31.03/0.8676\\ 
&TSRNet (Ours)&\textcolor{red}{37.92}/\textcolor{red}{0.9607}&\textcolor{red}{34.22}/\textcolor{red}{0.9261}&31.94/\textcolor{red}{0.8915}\\
\hline 
\end{tabular}
\end{center}
\end{table}

\begin{table}
\begin{center}
\caption{PSNR and SSIM results for different methods with different scales on Set14.}
\label{tab3}
\begin{tabular}{>{\centering\arraybackslash}p{0.8cm} >{\centering\arraybackslash}p{1.8cm} >{\centering\arraybackslash}p{1.24cm} >{\centering\arraybackslash}p{1.24cm} >{\centering\arraybackslash}p{1.24cm}}
\hline
\multirow{3}{*}{Dataset} & \multirow{3}{*}{Methods} 
& $\times 2 $ & $\times 3 $ & $\times 4 $ \\\cline{3-5}
& & PSNR(dB)& PSNR(dB) & PSNR(dB) \\
& & /SSIM & /SSIM & /SSIM \\\hline
\multirow{33}{*}{Set14} 
&Bicubic&30.24/0.8688&27.55/0.7742 &26.00/0.7027\\ 
&SRCNN\cite{dong2014learning}&32.42/0.9063&29.28/0.8209 &27.49/0.7503 \\
&VDSR\cite{kim2016accurate}&33.03/0.9124&29.77/0.8314 &28.01/0.7674\\
&DRRN\cite{tai2017image}&33.23/0.9136&29.96/0.8349 &28.21/0.7720\\
&FSRCNN\cite{dong2016accelerating}&32.63/0.9088&29.43/0.8242 &27.59/0.7535\\
&CARN-M\cite{ahn2018fast}&33.26/0.9141 &30.08/0.8367 &28.42/0.7762\\
&IDN\cite{hui2018fast}&33.30/0.9148 &29.99/0.8354 &28.25/0.7730\\
&A+\cite{timofte2015a+}&32.28/0.9056 &29.13/0.8188 &27.32/0.7491\\
&JOR\cite{dai2015jointly}&32.38/0.9063 &29.19/0.8204 &27.27/0.7479\\
&RFL\cite{schulter2015fast}&32.26/0.9040 &29.05/0.8164 &27.24/0.7451\\
&SelfEx\cite{huang2015single}&32.22/0.9034 &29.16/0.8196 &27.40/0.7518\\
&CSCN\cite{wang2015deep}&32.56/0.9074 &29.41/0.8238 &27.64/0.7578\\
&RED\cite{mao2016image}&32.81/0.9135 &29.50/0.8334 &27.72/0.7698\\ 
&MAP\cite{zhang2012single}&33.03/0.9128 &29.81/0.8321 &28.04/0.7672\\
&TNRD\cite{chen2016trainable}&32.51/0.9069 &29.43/0.8232 &27.66/0.7563\\ 
&FDSR\cite{lu2018fast}]&33.00/0.9042 &29.61/0.8179 &27.86/0.7500\\
&RCN\cite{shi2017structure}&32.77/0.9109 &29.63/0.8269 &27.79/0.7594\\
&DRCN\cite{kim2016deeply}&33.04/0.9118 &29.76/0.8311 &28.02/0.7670\\
&CNF\cite{ren2017image}&33.38/0.9136 &29.90/0.8322 &28.15/0.7680\\
&LapSRN\cite{lai2017deep}&33.08/0.9130 &29.63/0.8269&28.19/0.7720\\
&WaveResNet\cite{bae2017beyond}&33.09/0.9129 &29.88/0.8331&28.11/0.7699\\
&CPCA\cite{xu2018self}&31.04/0.8951&27.89/0.8038&26.10/0.7296\\
&NDRCN\cite{cao2019new}&33.20/0.9141&29.88/0.8333&28.10/0.7697\\\
&MemNet\cite{tai2017memnet}&33.28/0.9142 &30.00/0.8350 &28.26/0.7723\\
&LESRCNN\cite{tian2020lightweight}&33.32/0.9148 &30.12/0.8380 &28.44/0.7772\\
&LESRCNN-S\cite{tian2020lightweight}&33.30/0.9145&30.16/0.8384&28.43/0.7776\\
&DSRCNN\cite{song2022dual}&\textcolor{blue}{33.43}/\textcolor{blue}{0.9157} &\textcolor{blue}{30.24}/\textcolor{blue}{0.8402}&\textcolor{blue}{28.46}/\textcolor{red}{0.7796}\\
&DAN\cite{huang2020unfolding}&33.08/0.9041&30.09/0.8287&28.42/0.7687\\ 
&AFAN-S\cite{wang2023image}&33.40/0.9165&30.21/0.8405&28.47/0.7788\\
&CRFAN\cite{liu2023cross}&33.42/0.9170&-&8.35/0.7790\\
&ACDMSR\cite{niu2024acdmsr}&32.28/0.8863&28.76/0.7953&27.04/0.7341\\
&SAM-DiffSR\cite{wang2024sam}&-&-&27.14/0.7484\\
&TSRNet (Ours)&\textcolor{red}{33.59}/\textcolor{red}{0.9178}&\textcolor{red}{30.30}/\textcolor{red}{0.8417}&\textcolor{red}{28.50}/\textcolor{blue}{0.7791}\\
\hline 
\end{tabular}
\end{center}
\end{table}

\begin{table}
\begin{center}
\caption{PSNR and SSIM results for different methods with different scales on the B100.}
\label{tab4}
\begin{tabular}{>{\centering\arraybackslash}p{0.8cm} >{\centering\arraybackslash}p{1.8cm} >{\centering\arraybackslash}p{1.24cm} >{\centering\arraybackslash}p{1.24cm} >{\centering\arraybackslash}p{1.24cm}}
\hline
\multirow{3}{*}{Dataset} & \multirow{3}{*}{Methods} 
& $\times 2 $ & $\times 3 $ & $\times 4 $ \\\cline{3-5}
& & PSNR(dB)& PSNR(dB) & PSNR(dB) \\
& & /SSIM & /SSIM & /SSIM \\\hline
\multirow{33}{*}{B100} 
&Bicubic&29.56/0.8431&27.21/0.7385&25.96/0.6675\\
&SRCNN\cite{dong2014learning}&31.36/0.8879 &28.41/0.7863 &26.90/0.7101 \\
&VDSR\cite{kim2016accurate}&31.90/0.8960 &28.82/0.7976 &27.29/0.7251\\
&DRRN\cite{tai2017image}&32.05/0.8973 &28.95/0.8004 &27.38/0.7284 \\
&FSRCNN\cite{dong2016accelerating}&31.53/0.8920 &28.53/0.7910 &26.98/0.7150\\
&CARN-M\cite{ahn2018fast}&31.92/0.8960 &28.91/0.8000 &27.44/0.7304 \\
&IDN\cite{hui2018fast}&32.08/0.8985 &28.95/0.8013 &27.41/0.7297\\
&A+\cite{timofte2015a+}&31.21/0.8863 &28.29/0.7835&	26.82/0.7087\\
&JOR\cite{dai2015jointly}&31.22/0.8867 &28.27/0.7837 &26.79/0.7083\\
&RFL\cite{schulter2015fast}&31.16/0.8840 &28.22/0.7806 &26.75/0.7054\\
&SelfEx\cite{huang2015single}&31.18/0.8855 &28.29/0.7840 &26.84/0.7106\\
&CSCN\cite{wang2015deep}&31.40/0.8884 &28.50/0.7885 &27.03/0.7161 \\
&RED\cite{mao2016image}&31.96/0.8972 &28.88/0.7993 &27.35/0.7276\\
&MAP\cite{zhang2012single}&31.90/0.8961 &28.85/0.7981 &27.29/0.7253\\
&TNRD\cite{chen2016trainable}&31.40/0.8878 &28.50/0.7881 &27.00/0.7140\\ 
&FDSR\cite{lu2018fast}&31.87/0.8847 &28.82/0.7797 &27.31/0.7031 \\
&DRCN\cite{kim2016deeply}&31.85/0.8942 &28.80/0.7963 &27.23/0.7233\\
&CNF\cite{ren2017image}&31.91/0.8962 &28.82/0.7980 &27.32/0.7253\\
&LapSRN\cite{lai2017deep}&31.80/0.8950&-&27.32/0.7280\\
&NDRCN\cite{cao2019new}&32.00/0.8975&28.86/0.7991&27.30/0.7263\\
&MemNet\cite{tai2017memnet}&\textcolor{blue}{32.08}/0.8978 &28.96/0.8001&27.40/0.7281\\
&LESRCNN\cite{tian2020lightweight}&31.95/0.8964 &28.91/0.8005 &27.45/0.7313 \\
&LESRCNN-S\cite{tian2020lightweight}&31.95/0.8965&28.94/ 0.8012&27.47/0.7321\\\
&DSRCNN\cite{song2022dual}&32.05/0.8978 &\textcolor{blue}{29.01}/0.8029&	27.50/0.7341\\
&DAN\cite{huang2020unfolding}&31.76/0.8858&28.94/0.7919&\textcolor{blue}{27.51}/0.7248\\
&PGAN\cite{shi2023structure}&-&-&26.35/0.6926\\
&AFAN-S\cite{wang2023image}&32.05/\textcolor{blue}{0.8985}&28.99/\textcolor{blue}{0.8032}&27.49/0.7340\\
&ACDMSR\cite{niu2024acdmsr}&30.35/0.8588&27.16/0.7417&25.95/0.6743\\
&SAM-DiffSR\cite{wang2024sam}&-&-&25.54/\textcolor{blue}{0.7721}\\
&TSRNet (Ours)&\textcolor{red}{32.15}/\textcolor{red}{0.8999}&\textcolor{red}{29.06}/\textcolor{red}{0.8052}&\textcolor{red}{27.53}/\textcolor{red}{0.7353}\\
\hline 
\end{tabular}
\end{center}
\end{table}

\begin{table}
\begin{center}
\caption{PSNR and SSIM results for different methods with different scale on U100.}
\label{tab5}
\begin{tabular}{>{\centering\arraybackslash}p{0.8cm} >{\centering\arraybackslash}p{1.8cm} >{\centering\arraybackslash}p{1.24cm} >{\centering\arraybackslash}p{1.24cm} >{\centering\arraybackslash}p{1.24cm}}
\hline
\multirow{3}{*}{Dataset} & \multirow{3}{*}{Methods} 
& $\times 2 $ & $\times 3 $ & $\times 4 $ \\\cline{3-5}
& & PSNR(dB)& PSNR(dB) & PSNR(dB) \\
& & /SSIM & /SSIM & /SSIM \\\hline
\multirow{33}{*}{U100} 
&Bicubic&26.88/0.8403&24.46/0.7349&23.14/0.6577\\
&SRCNN\cite{dong2014learning}&29.50/0.8946&26.24/0.7989 &24.52/0.7221\\
&VDSR\cite{kim2016accurate}&30.76/0.9140&27.14/0.8279 &25.18/0.7524 \\
&DRRN\cite{tai2017image}&31.23/0.9188&27.53/0.8378 &25.44/0.7638 \\
&FSRCNN\cite{dong2016accelerating}&29.88/0.9020&26.43/0.8080 &24.62/0.7280\\
&CARN-M\cite{ahn2018fast}&31.23/0.9193&27.55/0.8385 &25.62/0.7694\\
&IDN\cite{hui2018fast}&31.27/0.9196&27.42/0.8359 &25.41/0.7632 \\
&A+\cite{timofte2015a+}&29.20/0.8938&26.03/0.7973&24.32/0.7183\\
&JOR\cite{dai2015jointly}&29.25/0.8951 &25.97/0.7972 &24.29/0.7181 \\ 
&RFL\cite{schulter2015fast}&29.11/0.8904 &25.86/0.7900 &24.19/0.7096\\
&SelfEx\cite{huang2015single}&29.54/0.8967 &26.44/0.8088&24.79/0.7374\\
&RED\cite{mao2016image}&30.91/0.9159 &27.31/0.8303 &25.35/0.7587 \\
&MAP\cite{zhang2012single}&30.74/0.9139 &27.15/0.8276 &25.20/0.7521 \\ 
&TNRD\cite{chen2016trainable}&29.70/0.8994 &26.42/0.8076 &24.61/0.7291\\ 
&FDSR\cite{lu2018fast}&30.91/0.9088 &27.23/0.8190 &25.27/0.7417\\
&LapSRN\cite{lai2017deep}&30.41/0.9100&-&25.21/0.7560\\ 
&WaveResNet\cite{bae2017beyond}&30.96/0.9169&27.28/0.8334&25.36/0.7614\\
&CPCA\cite{xu2018self}&28.17/0.8990&25.61/0.8123&23.62/0.7257\\
&NDRCN\cite{cao2019new}&31.06/0.9175&27.23/0.8312&25.16/0.7546\\
&DRCN\cite{kim2016deeply}&30.75/0.9133 &27.15/0.8276 &25.14/0.7510\\ 
&MemNet\cite{tai2017memnet}&31.31/0.9195 &27.56/0.8376&	25.50/0.7630\\
&LESRCNN\cite{tian2020lightweight}&31.45/0.9206 &27.70/0.8415 &25.77/0.7732\\
&LESRCNN-S\cite{tian2020lightweight}&31.45/0.9207&27.76/0.8424&25.78/0.7739\\
&DSRCNN\cite{song2022dual}&\textcolor{blue}{31.83}/\textcolor{blue}{0.9252}&\textcolor{blue}{27.99}/\textcolor{blue}{0.8483}&\textcolor{blue}{25.94}/\textcolor{blue}{0.7815}\\
&DAN\cite{huang2020unfolding}&30.60/0.9060&27.65/0.8352&25.86/0.7721\\
&PGAN\cite{shi2023structure}&-&-&25.91/0.7786\\ 
&AFAN-S\cite{wang2023image}&31.70/0.9240&27.85/0.8461&25.82/0.7770\\
&ACDMSR\cite{niu2024acdmsr}&31.72/0.9152&27.89/0.8412&25.85/0.7792\\ 
&TSRNet (Ours)&\textcolor{red}{32.06}/\textcolor{red}{0.9284}&\textcolor{red}{28.08}/\textcolor{red}{0.8509}&\textcolor{red}{26.00}/\textcolor{red}{0.7826}\\
\hline 
\end{tabular}
\end{center}
\end{table}

\begin{figure}[!t]
\centering
\includegraphics[width=2.8in]{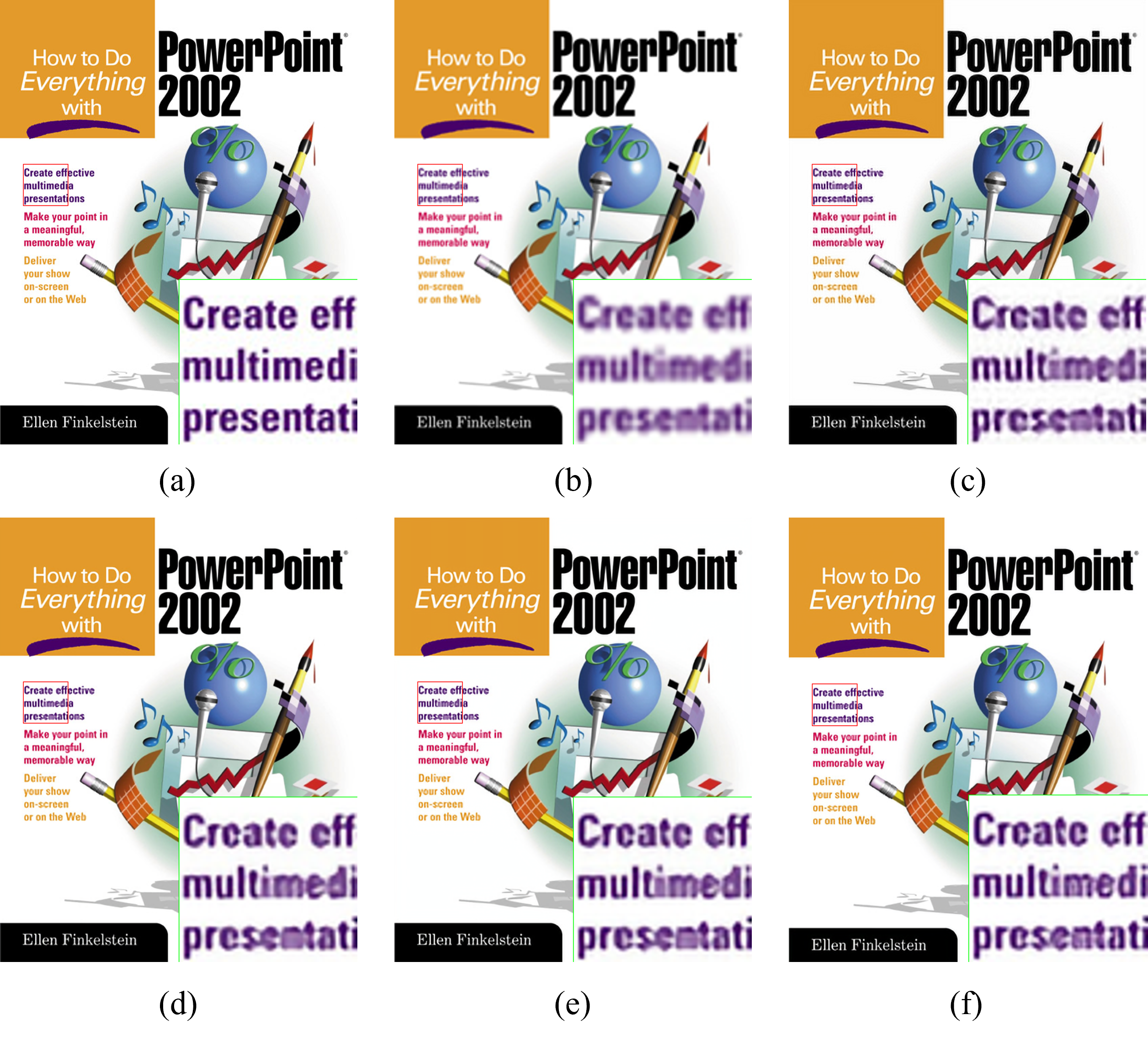}
\caption{Comparisons of visual results of different super-resolution methods on an image of  Set14 for $\times 2$. (a) HR image, (b) Bicubic, (c) SRCNN, (d) CARN-M, (e) LESRCNN and (f) TSRNet (Ours).}
\label{fig2}
\end{figure}
\begin{figure}[!t]
\centering
\includegraphics[width=2.8in]{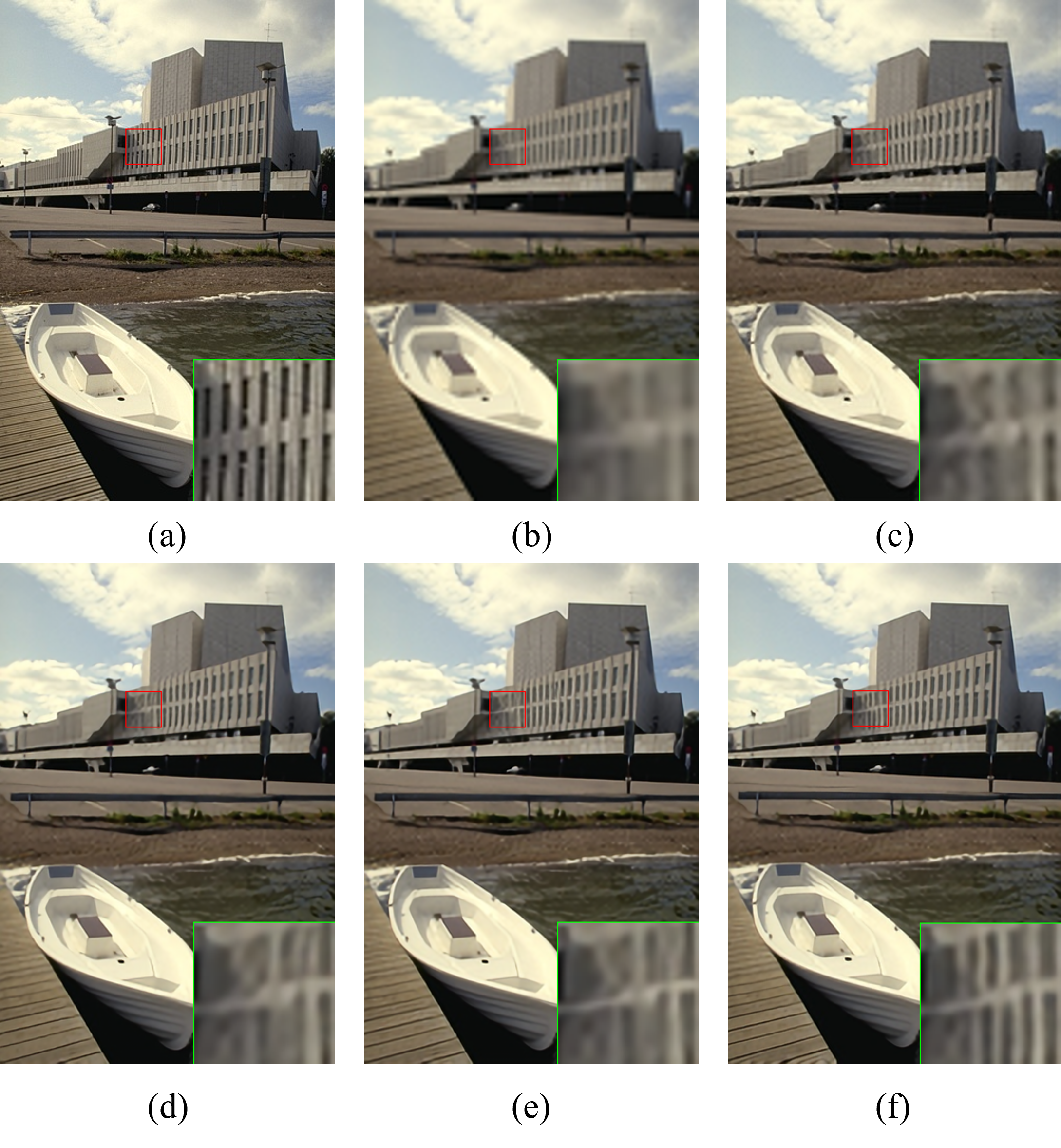}
\caption{Comparisons of visual results of different super-resolution methods on an image of  B100 for $\times 3$. (a) HR image, (b) Bicubic, (c) SRCNN, (d) CARN-M, (e) LESRCNN and (f) TSRNet (Ours).}
\label{fig3}
\end{figure}
\begin{figure}[!t]
\centering
\includegraphics[width=2.8in]{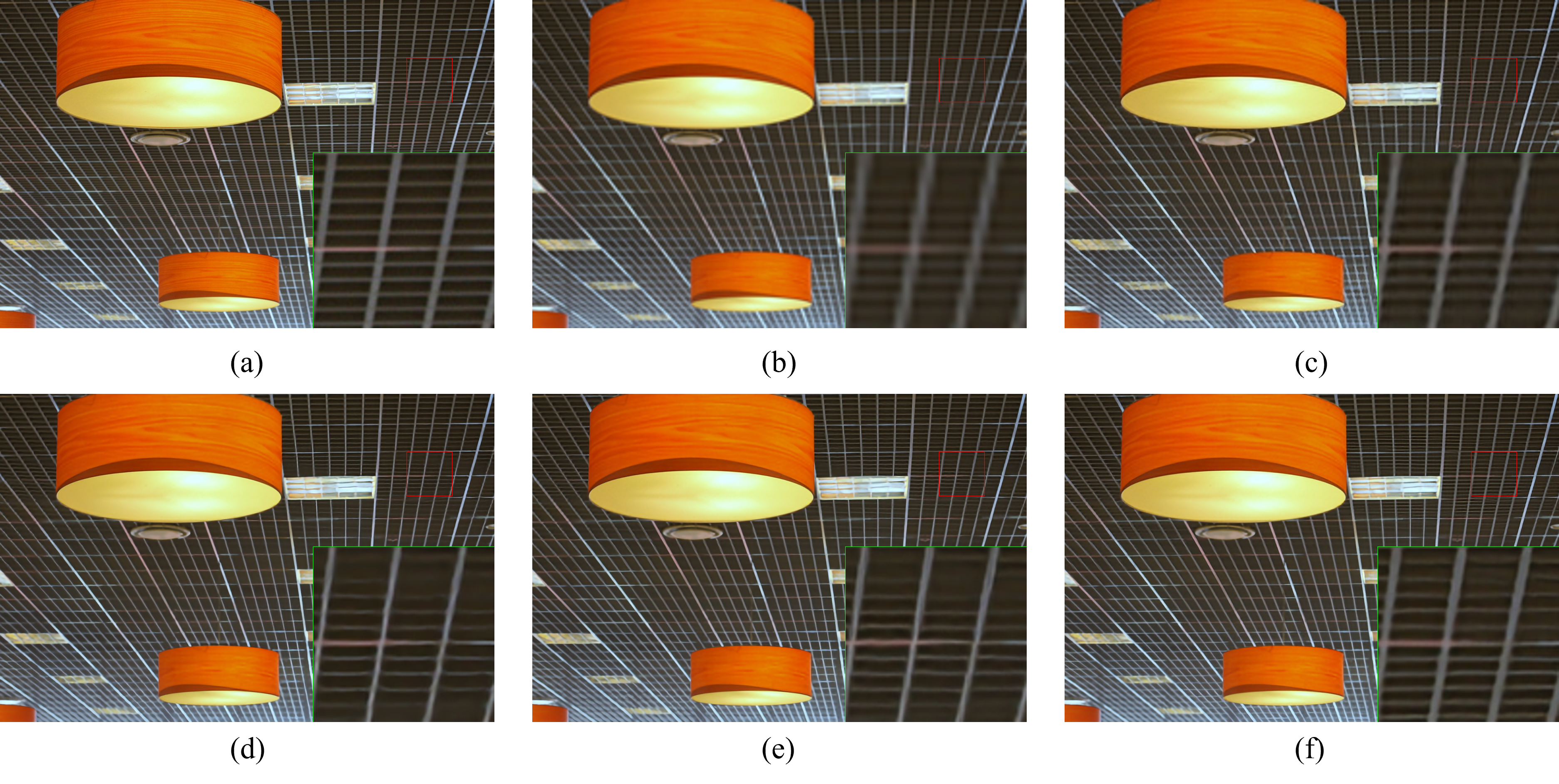}
\caption{Comparisons of visual results of different super-resolution methods on an image of  U100 for $\times 4$. (a) HR image, (b) Bicubic, (c) SRCNN, (d) CARN-M, (e) LESRCNN and (f) TSRNet (Ours).}
\label{fig4}
\end{figure}
\begin{figure}[!t]
\centering
\includegraphics[width=2.8in]{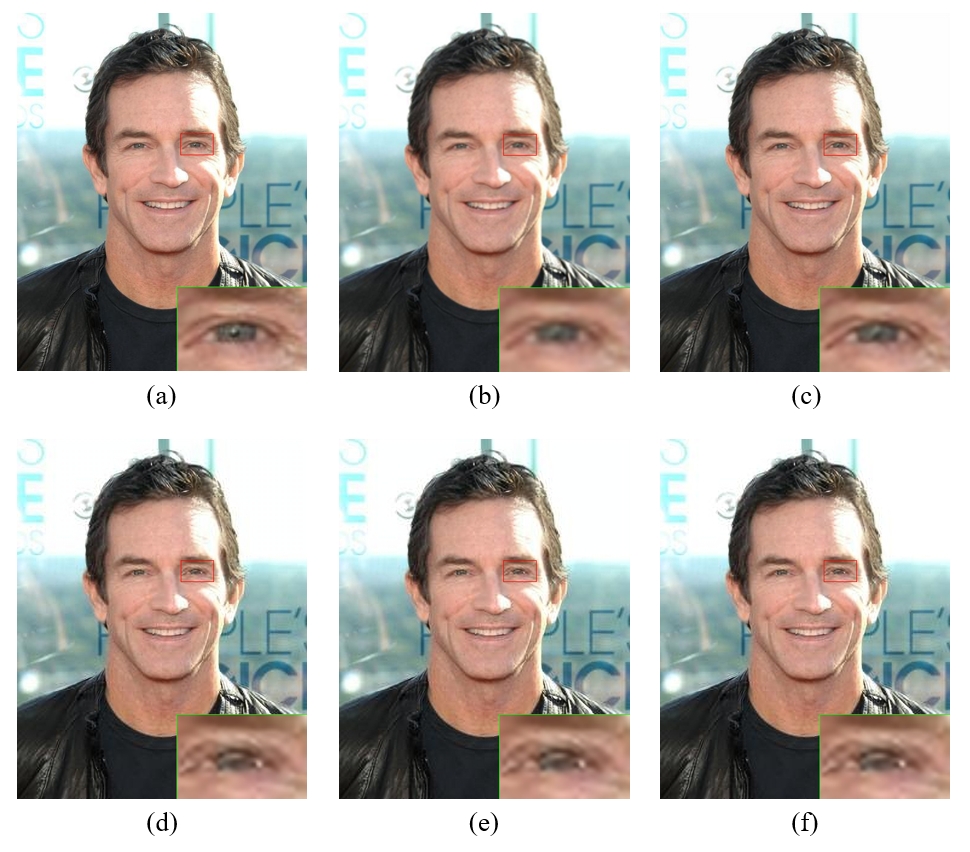}
\caption{Comparisons of visual results of different super-resolution methods on an image of  CelebA for $\times 2$. (a) HR image, (b) Bicubic, (c) SRCNN, (d) CARN-M, (e) LESRCNN and (f) TSRNet (Ours).}
\label{fig5}
\end{figure}
\begin{figure}[!t]
\centering
\includegraphics[width=2.8in]{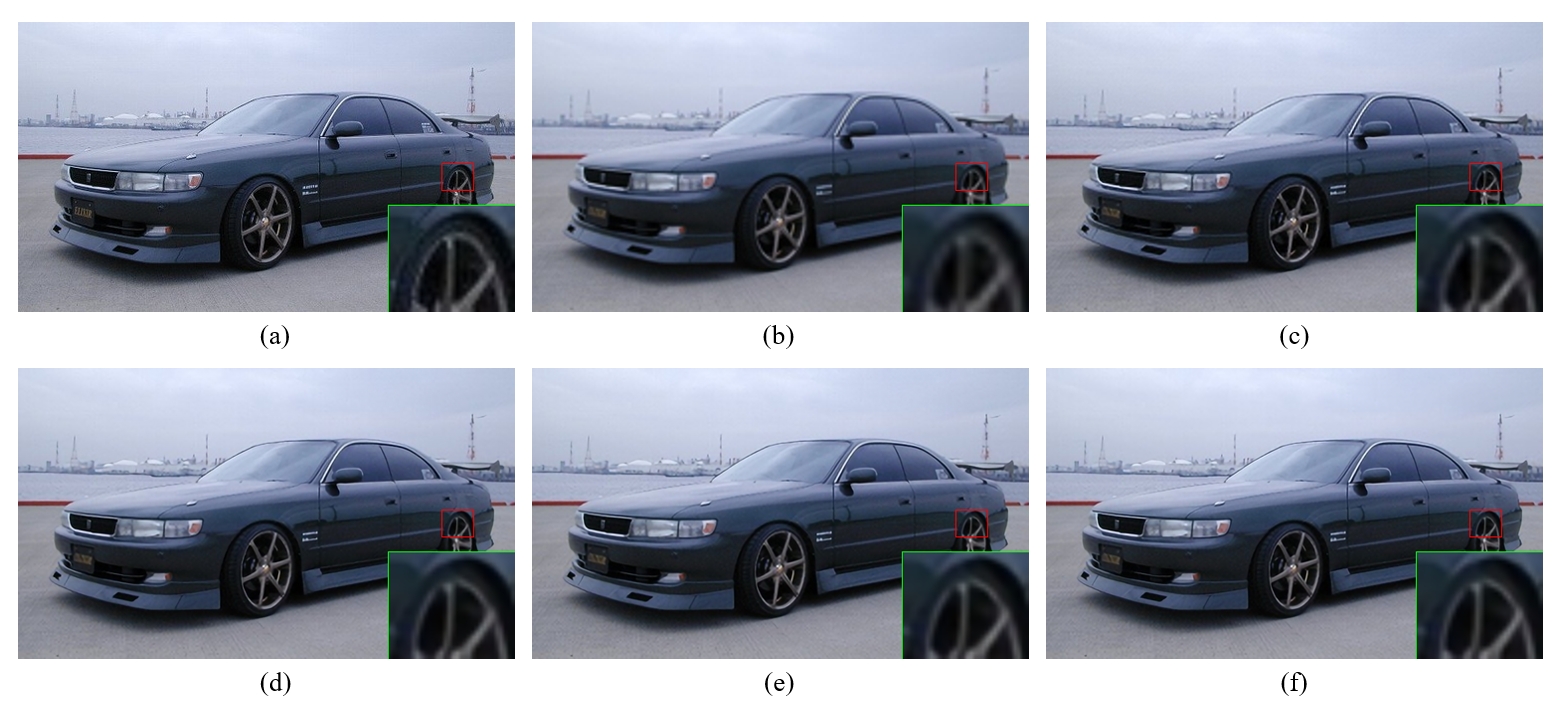}
\caption{Comparisons of visual results of different super-resolution methods on an image of  VOC2012 for $\times 2$. (a) HR image, (b) Bicubic, (c) SRCNN, (d) CARN-M, (e) LESRCNN and (f) TSRNet (Ours).}
\label{fig6}
\end{figure}
\subsection{Experimental results}
To fully evaluate image super-resolution performance of TSRNet, we analyze experimental results in both quantitative and qualitative analysis. Quantitative analysis is that uses  Bicubic method, SRCNN \cite{dong2014learning}, VDSR \cite{kim2016accurate}, DRRN \cite{tai2017image}, FSRCNN \cite{dong2016accelerating}, an accurate and lightweight deep network (CARN-M) \cite{ahn2018fast}, an information distillation network (IDN) \cite{hui2018fast}, adjusted anchored neighborhood regression (A+) \cite{timofte2015a+}, a jointly optimized regressors (JOR) \cite{dai2015jointly}, a random forest linear method (RFL) \cite{schulter2015fast}, a transformed self-exemplar (SelfEx) \cite{huang2015single}, a cascade of sparse coding based networks (CSCN) \cite{wang2015deep}], a very deep residual encoder-decoder network (RED) \cite{mao2016image}, a maximum a posteriori probability framework (MAP) \cite{zhang2012single}, a trainable nonlinear reaction diffusion network (TNRD) \cite{chen2016trainable}, a symmetrical dilated residual convolution network (FDSR) \cite{lu2018fast}, a residue context sub-network (RCN) \cite{shi2017structure}, DRCN \cite{kim2016deeply}, a context-wise network fusion approach (CNF) \cite{ren2017image}, a Laplacian pyramid super-resolution network (LapSRN) \cite{lai2017deep}, a wavelet domain residual network (WaveResNet) \cite{bae2017beyond}, convolutional principal component analysis (CPCA) \cite{xu2018self}, a new architecture of deep recursive convolution network (NDRCN) \cite{cao2019new}, a persistent memory network (MemNet) \cite{tai2017memnet}, a lightweight enhanced super-resolution CNN (LESRCNN) \cite{tian2020lightweight}, LESRCNN-S \cite{tian2020lightweight}, a dual super-resolution CNN (DSRCNN) \cite{song2022dual}, a deep alternating network (DAN) \cite{huang2020unfolding}, deep constrained least squares (DCLS) \cite{luo2022deep}], a pixel-Level generative adversarial network (PGAN) \cite{shi2023structure}, ACNet\cite{tian2021asymmetric}, an attention-directed feature aggregation network (AFAN-S) \cite{wang2023image}, a cross-resolution attention network (CRFAN) \cite{liu2023cross},  an accelerated conditional diffusion model (ACDMSR) \cite{niu2024acdmsr}, a structure-modulated diffusion model (SAM-DiffSR) \cite{wang2024sam} , an enhanced deep residual network (EDSR)\cite{lim2017enhanced} as comparative methods on four public datasets, i.e., Set5, Set14, B100 and U100 for $\times 2$, $\times 3$ and $\times 4$ to evaluate SR performance of our TSRNet in terms of peak signal-to-noise ration (PSNR) \cite{hore2010image} and structural similarity index (SSIM) \cite{wang2004image}. For small datasets, we can see that our TSRNet is superior than that of the second method in TABLEs \ref{tab2} and \ref{tab3}.  For instance, our TSRNet has exceeded 0.08dB of PSNR than that of AFAN-S for $\times 2$ in TABLE \ref{tab2}. Our TSRNet has obtained an improvement of 0.06dB of PSNR and 0.0015 of SSIM than that of  DSRCNN for $\times 3$ in TABLE \ref{tab3}. For big dataset, the proposed TSRNet has obtained excellent SR results. As shown in TABLE \ref{tab4}, we can see that our TSRNet has an improvement of 0.07dB of PSNR for $\times 2$ than that of the second MemNet on B100.  As illustrated in TABLE \ref{tab5}, our TSRNet has an improvement of 0.09dB than that of DSRCNN for $\times 3$ on U100. According to mentioned illustrations, we can see that the proposed method is superior to popular SR methods for four public datasets. Besides, our method is superior to running time in TABLE \ref{tab6} and complexity in TABLE \ref{tab7}. Thus, our method is effective for image super-resolution in quantitative analysis.

In terms of qualitative analysis, to evaluate the visual quality of suer-resolution images, we compare the predicted super-resolution images from different methods i.e. Bicubic, VDSR, CARN-M, LESRCNN and TSRNet to observe the visual effect. Specifically, we select three images from Set14, B100 and U100 on $\times 2$, $\times 3$ and $\times 4$ scales as shown in Figures \ref{fig2}, \ref{fig3} and \ref{fig4}, respectively. It can be seen that visual images of the TSRNet are clearer in details. To obtain the conclusion of model generalization, we test on the face dataset CelebA\cite{liu2015deep} and image recognition dataset VOC2012\cite{everingham2015pascal}, respectively. We subsample the original images of the two datasets to simulate low-quality images. And the original images are regarded as high-quality images. The visual comparison results with other models are shown in Fig. \ref{fig5} and \ref{fig6}. As shown in Fig. \ref{fig5}, our model can effectively restore wrinkles in the corners of the face and eyes. As shown in Fig. \ref{fig6}, the vehicle hub restored by our model is clearer. The experimental results show that our network can achieve better visual results. According to mentioned illustrations, we can see that our TSRNet can better deal with recovering high-quality images in quantitative and qualitative analysis.

\begin{table}
\begin{center}
\caption{Running time (ms ) of different methods with $\times 4$.}
\label{tab6}
\begin{tabular}{>{\centering\arraybackslash}p{1.6cm} >{\centering\arraybackslash}p{1.2cm} >{\centering\arraybackslash}p{1.2cm} >{\centering\arraybackslash}p{1.2cm} >{\centering\arraybackslash}p{1.2cm}}
\hline
Images sizes & EDSR\cite{lim2017enhanced} & CARN-M\cite{ahn2018fast} & ACNet\cite{tian2021asymmetric} & TSRNet (Ours)\\\hline
256 × 256&17.22&15.91&18.38&13.34\\
512 × 512&57.53&45.21&75.79&35.68\\\hline
\end{tabular}
\end{center}
\end{table}
\begin{table}
\begin{center}
\caption{Parameters and FLOPs of different methods with $\times 4$ for image SR with $1024 \times 1024$.}
\label{tab7}
\begin{tabular}{ c  c  c }
\hline
Methods&Parameters&FLOPs\\\hline
DCLS\cite{luo2022deep}&13626K&498.18G\\
TSRNet (Ours)&2253K&181.25G\\
\hline
\end{tabular}
\end{center}
\end{table}
\section{Conclusion}
In this paper, we propose a tree-guided network for image super-resolution. It can use relations of different trees to improve effects of hierarchical information to advance SR effect. To prevent loss of local key information, cosine transform techniques are used to extract similarity of direction features to obtain local salient information for image SR. To prevent local optimum, Adan is employed to reduce gradient explosion to optimize parameters to further enhance the quality of the predicted images. Quantitative and qualitative analysis shows that our proposed method is superior to other popular SR methods. In the future, we will study the relations of different trees for image super-resolution. 

\bibliographystyle{IEEEtran}

\begin{thebibliography}{10}
\providecommand{\url}[1]{#1}
\csname url@samestyle\endcsname
\providecommand{\newblock}{\relax}
\providecommand{\bibinfo}[2]{#2}
\providecommand{\BIBentrySTDinterwordspacing}{\spaceskip=0pt\relax}
\providecommand{\BIBentryALTinterwordstretchfactor}{4}
\providecommand{\BIBentryALTinterwordspacing}{\spaceskip=\fontdimen2\font plus
\BIBentryALTinterwordstretchfactor\fontdimen3\font minus
  \fontdimen4\font\relax}
\providecommand{\BIBforeignlanguage}[2]{{%
\expandafter\ifx\csname l@#1\endcsname\relax
\typeout{** WARNING: IEEEtran.bst: No hyphenation pattern has been}%
\typeout{** loaded for the language `#1'. Using the pattern for}%
\typeout{** the default language instead.}%
\else
\language=\csname l@#1\endcsname
\fi
#2}}
\providecommand{\BIBdecl}{\relax}
\BIBdecl

\bibitem{yang2019deep}
W.~Yang, X.~Zhang, Y.~Tian, W.~Wang, J.-H. Xue, and Q.~Liao, ``Deep learning
  for single image super-resolution: A brief review,'' \emph{IEEE Transactions
  on Multimedia}, vol.~21, no.~12, pp. 3106--3121, 2019.

\bibitem{rukundo2012nearest}
O.~Rukundo and H.~Cao, ``Nearest neighbor value interpolation,'' \emph{arXiv
  preprint arXiv:1211.1768}, 2012.

\bibitem{li2001new}
X.~Li and M.~T. Orchard, ``New edge-directed interpolation,'' \emph{IEEE
  transactions on image processing}, vol.~10, no.~10, pp. 1521--1527, 2001.

\bibitem{keys1981cubic}
R.~Keys, ``Cubic convolution interpolation for digital image processing,''
  \emph{IEEE transactions on acoustics, speech, and signal processing},
  vol.~29, no.~6, pp. 1153--1160, 1981.

\bibitem{park2003super}
S.~C. Park, M.~K. Park, and M.~G. Kang, ``Super-resolution image
  reconstruction: a technical overview,'' \emph{IEEE signal processing
  magazine}, vol.~20, no.~3, pp. 21--36, 2003.

\bibitem{liao2023minimax}
X.~Liao, X.~Wei, and M.~Zhou, ``Minimax concave penalty regression for
  superresolution image reconstruction,'' \emph{IEEE Transactions on Consumer
  Electronics}, vol.~70, no.~1, pp. 2999-3007, 2023.

\bibitem{yang2010image}
J.~Yang, J.~Wright, T.~S. Huang, and Y.~Ma, ``Image super-resolution via sparse
  representation,'' \emph{IEEE transactions on image processing}, vol.~19,
  no.~11, pp. 2861--2873, 2010.

\bibitem{yang2008image}
J.~Yang, J.~Wright, T.~Huang, and Y.~Ma, ``Image super-resolution as sparse
  representation of raw image patches,'' in \emph{2008 IEEE conference on
  computer vision and pattern recognition}.\hskip 1em plus 0.5em minus
  0.4em\relax IEEE, 2008, pp. 1--8.

\bibitem{rousseau2010non}
F.~Rousseau, A.~D.~N. Initiative \emph{et~al.}, ``A non-local approach for
  image super-resolution using intermodality priors,'' \emph{Medical image
  analysis}, vol.~14, no.~4, pp. 594--605, 2010.

\bibitem{dong2014learning}
C.~Dong, C.~C. Loy, K.~He, and X.~Tang, ``Learning a deep convolutional network
  for image super-resolution,'' in \emph{Computer Vision--ECCV 2014: 13th
  European Conference, Zurich, Switzerland, September 6-12, 2014, Proceedings,
  Part IV 13}.\hskip 1em plus 0.5em minus 0.4em\relax Springer, 2014, pp.
  184--199.

\bibitem{tian2020coarse}
C.~Tian, Y.~Xu, W.~Zuo, B.~Zhang, L.~Fei, and C.-W. Lin, ``Coarse-to-fine cnn
  for image super-resolution,'' \emph{IEEE Transactions on Multimedia},
  vol.~23, pp. 1489--1502, 2020.

\bibitem{dong2015image}
C.~Dong, C.~C. Loy, K.~He, and X.~Tang, ``Image super-resolution using deep
  convolutional networks,'' \emph{IEEE transactions on pattern analysis and
  machine intelligence}, vol.~38, no.~2, pp. 295--307, 2015.

\bibitem{hu2019runet}
X.~Hu, M.~A. Naiel, A.~Wong, M.~Lamm, and P.~Fieguth, ``Runet: A robust unet
  architecture for image super-resolution,'' in \emph{Proceedings of the
  IEEE/CVF Conference on Computer Vision and Pattern Recognition Workshops},
  2019, pp. 0--0.

\bibitem{kim2016accurate}
J.~Kim, J.~K. Lee, and K.~M. Lee, ``Accurate image super-resolution using very
  deep convolutional networks,'' in \emph{Proceedings of the IEEE conference on
  computer vision and pattern recognition}, 2016, pp. 1646--1654.

\bibitem{kim2016deeply}
J.~Kim, J.~K. Lee, and K.~M. Lee, ``Deeply-recursive convolutional network for image super-resolution,''
  in \emph{Proceedings of the IEEE conference on computer vision and pattern
  recognition}, 2016, pp. 1637--1645.

\bibitem{dong2016accelerating}
C.~Dong, C.~C. Loy, and X.~Tang, ``Accelerating the super-resolution
  convolutional neural network,'' in \emph{Computer Vision--ECCV 2016: 14th
  European Conference, Amsterdam, The Netherlands, October 11-14, 2016,
  Proceedings, Part II 14}.\hskip 1em plus 0.5em minus 0.4em\relax Springer,
  2016, pp. 391--407.

\bibitem{tian2021asymmetric}
C.~Tian, Y.~Xu, W.~Zuo, C.-W. Lin, and D.~Zhang, ``Asymmetric cnn for image
  superresolution,'' \emph{IEEE Transactions on Systems, Man, and Cybernetics:
  Systems}, vol.~52, no.~6, pp. 3718--3730, 2021.

\bibitem{wanga2023hybrid}
K.~Wanga, X.~Yanga, and G.~Jeon, ``Hybrid attention feature refinement network
  for lightweight image super-resolution in metaverse immersive display,''
  \emph{IEEE Transactions on Consumer Electronics}, vol.~70, no.~1, pp. 3232--3244, 2023.

\bibitem{fan2018compressed}
X.~Fan, Y.~Yang, C.~Deng, J.~Xu, and X.~Gao, ``Compressed multi-scale feature
  fusion network for single image super-resolution,'' \emph{Signal processing},
  vol. 146, pp. 50--60, 2018.

\bibitem{tai2017memnet}
Y.~Tai, J.~Yang, X.~Liu, and C.~Xu, ``Memnet: A persistent memory network for
  image restoration,'' in \emph{Proceedings of the IEEE international
  conference on computer vision}, 2017, pp. 4539--4547.

\bibitem{tai2017image}
Y.~Tai, J.~Yang, and X.~Liu, ``Image super-resolution via deep recursive
  residual network,'' in \emph{Proceedings of the IEEE conference on computer
  vision and pattern recognition}, 2017, pp. 3147--3155.

\bibitem{wang2020deep}
Z.~Wang, J.~Chen, and S.~C. Hoi, ``Deep learning for image super-resolution: A
  survey,'' \emph{IEEE transactions on pattern analysis and machine
  intelligence}, vol.~43, no.~10, pp. 3365--3387, 2020.

\bibitem{tong2017image}
T.~Tong, G.~Li, X.~Liu, and Q.~Gao, ``Image super-resolution using dense skip
  connections,'' in \emph{Proceedings of the IEEE international conference on
  computer vision}, 2017, pp. 4799--4807.

\bibitem{hui2019lightweight}
Z.~Hui, X.~Gao, Y.~Yang, and X.~Wang, ``Lightweight image super-resolution with
  information multi-distillation network,'' in \emph{Proceedings of the 27th
  acm international conference on multimedia}, 2019, pp. 2024--2032.

\bibitem{ledig2017photo}
C.~Ledig, L.~Theis, F.~Husz{\'a}r, J.~Caballero, A.~Cunningham, A.~Acosta,
  A.~Aitken, A.~Tejani, J.~Totz, Z.~Wang \emph{et~al.}, ``Photo-realistic
  single image super-resolution using a generative adversarial network,'' in
  \emph{Proceedings of the IEEE conference on computer vision and pattern
  recognition}, 2017, pp. 4681--4690.

\bibitem{gao2023ctcnet}
G.~Gao, Z.~Xu, J.~Li, J.~Yang, T.~Zeng, and G.-J. Qi, ``Ctcnet: A
  cnn-transformer cooperation network for face image super-resolution,''
  \emph{IEEE Transactions on Image Processing}, vol.~32, pp. 1978--1991, 2023.

\bibitem{bulat2018super}
A.~Bulat and G.~Tzimiropoulos, ``Super-fan: Integrated facial landmark
  localization and super-resolution of real-world low resolution faces in
  arbitrary poses with gans,'' in \emph{Proceedings of the IEEE conference on
  computer vision and pattern recognition}, 2018, pp. 109--117.

\bibitem{lai2017deep}
W.-S. Lai, J.-B. Huang, N.~Ahuja, and M.-H. Yang, ``Deep laplacian pyramid
  networks for fast and accurate super-resolution,'' in \emph{Proceedings of
  the IEEE conference on computer vision and pattern recognition}, 2017, pp.
  624--632.

\bibitem{johnson2016perceptual}
J.~Johnson, A.~Alahi, and L.~Fei-Fei, ``Perceptual losses for real-time style
  transfer and super-resolution,'' in \emph{Computer Vision--ECCV 2016: 14th
  European Conference, Amsterdam, The Netherlands, October 11-14, 2016,
  Proceedings, Part II 14}.\hskip 1em plus 0.5em minus 0.4em\relax Springer,
  2016, pp. 694--711.

\bibitem{robbins1951stochastic}
H.~Robbins and S.~Monro, ``A stochastic approximation method,'' \emph{The
  annals of mathematical statistics}, pp. 400--407, 1951.

\bibitem{kingma2014adam}
D.~P. Kingma, ``Adam: A method for stochastic optimization,'' \emph{arXiv
  preprint arXiv:1412.6980}, 2014.

\bibitem{lim2017enhanced}
B.~Lim, S.~Son, H.~Kim, S.~Nah, and K.~Mu~Lee, ``Enhanced deep residual
  networks for single image super-resolution,'' in \emph{Proceedings of the
  IEEE conference on computer vision and pattern recognition workshops}, 2017,
  pp. 136--144.

\bibitem{loshchilov2017decoupled}
I.~Loshchilov, ``Decoupled weight decay regularization,'' \emph{arXiv preprint
  arXiv:1711.05101}, 2017.

\bibitem{xie2024adan}
X.~Xie, P.~Zhou, H.~Li, Z.~Lin, and S.~Yan, ``Adan: Adaptive nesterov momentum
  algorithm for faster optimizing deep models,'' \emph{IEEE Transactions on
  Pattern Analysis and Machine Intelligence}, 2024.

\bibitem{tian2022image}
C.~Tian, Y.~Yuan, S.~Zhang, C.-W. Lin, W.~Zuo, and D.~Zhang, ``Image
  super-resolution with an enhanced group convolutional neural network,''
  \emph{Neural Networks}, vol. 153, pp. 373--385, 2022.

\bibitem{pisoni2022sharpenedcossim}
R.~Pisoni, ``Sharpened cosine distance as an alternative for convolutions,''
  2022, \url{https://rpisoni.dev/posts/cossim-convolution/}.

\bibitem{agustsson2017ntire}
E.~Agustsson and R.~Timofte, ``Ntire 2017 challenge on single image
  super-resolution: Dataset and study,'' in \emph{Proceedings of the IEEE
  conference on computer vision and pattern recognition workshops}, 2017, pp.
  126--135.

\bibitem{bevilacqua2012low}
M.~Bevilacqua, A.~Roumy, C.~Guillemot, and M.~L. Alberi-Morel, ``Low-complexity
  single-image super-resolution based on nonnegative neighbor
  embedding.''\hskip 1em plus 0.5em minus 0.4em\relax BMVA press, 2012.

\bibitem{zeyde2012single}
R.~Zeyde, M.~Elad, and M.~Protter, ``On single image scale-up using
  sparse-representations,'' in \emph{Curves and Surfaces: 7th International
  Conference, Avignon, France, June 24-30, 2010, Revised Selected Papers
  7}.\hskip 1em plus 0.5em minus 0.4em\relax Springer, 2012, pp. 711--730.

\bibitem{martin2001database}
D.~Martin, C.~Fowlkes, D.~Tal, and J.~Malik, ``A database of human segmented
  natural images and its application to evaluating segmentation algorithms and
  measuring ecological statistics,'' in \emph{Proceedings eighth IEEE
  international conference on computer vision. ICCV 2001}, vol.~2.\hskip 1em
  plus 0.5em minus 0.4em\relax IEEE, 2001, pp. 416--423.

\bibitem{huang2015single}
J.-B. Huang, A.~Singh, and N.~Ahuja, ``Single image super-resolution from
  transformed self-exemplars,'' in \emph{Proceedings of the IEEE conference on
  computer vision and pattern recognition}, 2015, pp. 5197--5206.

\bibitem{wang2024multi}
Y.~Wang, Y.~Li, G.~Wang, and X.~Liu, ``Multi-scale attention network for single
  image super-resolution,'' in \emph{Proceedings of the IEEE/CVF Conference on
  Computer Vision and Pattern Recognition}, 2024, pp. 5950--5960.

\bibitem{zhang2018residual}
Y.~Zhang, Y.~Tian, Y.~Kong, B.~Zhong, and Y.~Fu, ``Residual dense network for
  image super-resolution,'' in \emph{Proceedings of the IEEE conference on
  computer vision and pattern recognition}, 2018, pp. 2472--2481.

\bibitem{yi2018separable}
S.~Yi and Y.~Zhou, ``Separable and reversible data hiding in encrypted images
  using parametric binary tree labeling,'' \emph{IEEE Transactions on
  Multimedia}, vol.~21, no.~1, pp. 51--64, 2018.

\bibitem{tian2022heterogeneous}
C.~Tian, Y.~Zhang, W.~Zuo, C.-W. Lin, D.~Zhang, and Y.~Yuan, ``A heterogeneous
  group cnn for image super-resolution,'' \emph{IEEE transactions on neural
  networks and learning systems}, 2022.

\bibitem{ahn2018fast}
N.~Ahn, B.~Kang, and K.-A. Sohn, ``Fast, accurate, and lightweight
  super-resolution with cascading residual network,'' in \emph{Proceedings of
  the European conference on computer vision (ECCV)}, 2018, pp. 252--268.

\bibitem{hui2018fast}
Z.~Hui, X.~Wang, and X.~Gao, ``Fast and accurate single image super-resolution
  via information distillation network,'' in \emph{Proceedings of the IEEE
  conference on computer vision and pattern recognition}, 2018, pp. 723--731.

\bibitem{timofte2015a+}
R.~Timofte, V.~De~Smet, and L.~Van~Gool, ``A+: Adjusted anchored neighborhood
  regression for fast super-resolution,'' in \emph{Computer Vision--ACCV 2014:
  12th Asian Conference on Computer Vision, Singapore, Singapore, November 1-5,
  2014, Revised Selected Papers, Part IV 12}.\hskip 1em plus 0.5em minus
  0.4em\relax Springer, 2015, pp. 111--126.

\bibitem{dai2015jointly}
D.~Dai, R.~Timofte, and L.~Van~Gool, ``Jointly optimized regressors for image
  super-resolution,'' in \emph{Computer Graphics Forum}, vol.~34, no.~2.\hskip
  1em plus 0.5em minus 0.4em\relax Wiley Online Library, 2015, pp. 95--104.

\bibitem{schulter2015fast}
S.~Schulter, C.~Leistner, and H.~Bischof, ``Fast and accurate image upscaling
  with super-resolution forests,'' in \emph{Proceedings of the IEEE conference
  on computer vision and pattern recognition}, 2015, pp. 3791--3799.

\bibitem{wang2015deep}
Z.~Wang, D.~Liu, J.~Yang, W.~Han, and T.~Huang, ``Deep networks for image
  super-resolution with sparse prior,'' in \emph{Proceedings of the IEEE
  international conference on computer vision}, 2015, pp. 370--378.

\bibitem{mao2016image}
X.~Mao, C.~Shen, and Y.-B. Yang, ``Image restoration using very deep
  convolutional encoder-decoder networks with symmetric skip connections,''
  \emph{Advances in neural information processing systems}, vol.~29, 2016.

\bibitem{zhang2012single}
K.~Zhang, X.~Gao, D.~Tao, and X.~Li, ``Single image super-resolution with
  non-local means and steering kernel regression,'' \emph{IEEE Transactions on
  Image Processing}, vol.~21, no.~11, pp. 4544--4556, 2012.

\bibitem{chen2016trainable}
Y.~Chen and T.~Pock, ``Trainable nonlinear reaction diffusion: A flexible
  framework for fast and effective image restoration,'' \emph{IEEE transactions
  on pattern analysis and machine intelligence}, vol.~39, no.~6, pp.
  1256--1272, 2016.

\bibitem{lu2018fast}
Z.~Lu, Z.~Yu, P.~Yali, L.~Shigang, W.~Xiaojun, L.~Gang, and R.~Yuan, ``Fast
  single image super-resolution via dilated residual networks,'' \emph{Ieee
  Access}, vol.~7, pp. 109\,729--109\,738, 2018.

\bibitem{shi2017structure}
Y.~Shi, K.~Wang, C.~Chen, L.~Xu, and L.~Lin, ``Structure-preserving image
  super-resolution via contextualized multitask learning,'' \emph{IEEE
  transactions on multimedia}, vol.~19, no.~12, pp. 2804--2815, 2017.

\bibitem{ren2017image}
H.~Ren, M.~El-Khamy, and J.~Lee, ``Image super resolution based on fusing
  multiple convolution neural networks,'' in \emph{Proceedings of the IEEE
  conference on computer vision and pattern recognition workshops}, 2017, pp.
  54--61.

\bibitem{bae2017beyond}
W.~Bae, J.~Yoo, and J.~Chul~Ye, ``Beyond deep residual learning for image
  restoration: Persistent homology-guided manifold simplification,'' in
  \emph{Proceedings of the IEEE conference on computer vision and pattern
  recognition workshops}, 2017, pp. 145--153.

\bibitem{xu2018self}
J.~Xu, M.~Li, J.~Fan, X.~Zhao, and Z.~Chang, ``Self-learning super-resolution
  using convolutional principal component analysis and random matching,''
  \emph{IEEE Transactions on Multimedia}, vol.~21, no.~5, pp. 1108--1121, 2018.

\bibitem{cao2019new}
F.~Cao and B.~Chen, ``New architecture of deep recursive convolution networks
  for super-resolution,'' \emph{Knowledge-Based Systems}, vol. 178, pp.
  98--110, 2019.

\bibitem{tian2020lightweight}
C.~Tian, R.~Zhuge, Z.~Wu, Y.~Xu, W.~Zuo, C.~Chen, and C.-W. Lin, ``Lightweight
  image super-resolution with enhanced cnn,'' \emph{Knowledge-Based Systems},
  vol. 205, p. 106235, 2020.

\bibitem{song2022dual}
J.~Song, J.~Xiao, C.~Tian, Y.~Hu, L.~You, and S.~Zhang, ``A dual cnn for image
  super-resolution,'' \emph{Electronics}, vol.~11, no.~5, p. 757, 2022.

\bibitem{huang2020unfolding}
Y.~Huang, S.~Li, L.~Wang, T.~Tan \emph{et~al.}, ``Unfolding the alternating
  optimization for blind super resolution,'' \emph{Advances in Neural
  Information Processing Systems}, vol.~33, pp. 5632--5643, 2020.

\bibitem{luo2022deep}
Z.~Luo, H.~Huang, L.~Yu, Y.~Li, H.~Fan, and S.~Liu, ``Deep constrained least
  squares for blind image super-resolution,'' in \emph{Proceedings of the
  IEEE/CVF conference on computer vision and pattern recognition}, 2022, pp.
  17\,642--17\,652.

\bibitem{shi2023structure}
W.~Shi, F.~Tao, and Y.~Wen, ``Structure-aware deep networks and pixel-level
  generative adversarial training for single image super-resolution,''
  \emph{IEEE Transactions on Instrumentation and Measurement}, vol.~72, pp.
  1--14, 2023.

\bibitem{wang2023image}
L.~Wang, K.~Li, J.~Tang, and Y.~Liang, ``Image super-resolution via lightweight
  attention-directed feature aggregation network,'' \emph{ACM Transactions on
  Multimedia Computing, Communications and Applications}, vol.~19, no.~2, pp.
  1--23, 2023.

\bibitem{niu2024acdmsr}
A.~Niu, T.~X. Pham, K.~Zhang, J.~Sun, Y.~Zhu, Q.~Yan, I.~S. Kweon, and
  Y.~Zhang, ``Acdmsr: Accelerated conditional diffusion models for single image
  super-resolution,'' \emph{IEEE Transactions on Broadcasting}, 2024.

\bibitem{liu2023cross}
A.~Liu, S.~Li, and Y.~Chang, ``Cross-resolution feature attention network for
  image super-resolution,'' \emph{The Visual Computer}, vol.~39, no.~9, pp.
  3837--3849, 2023.

\bibitem{wang2024sam}
C.~Wang, Z.~Hao, Y.~Tang, J.~Guo, Y.~Yang, K.~Han, and Y.~Wang, ``Sam-diffsr:
  Structure-modulated diffusion model for image super-resolution,'' \emph{arXiv
  preprint arXiv:2402.17133}, 2024.

\bibitem{hore2010image}
A.~Hore and D.~Ziou, ``Image quality metrics: Psnr vs. ssim,'' in \emph{2010
  20th international conference on pattern recognition}.\hskip 1em plus 0.5em
  minus 0.4em\relax IEEE, 2010, pp. 2366--2369.

\bibitem{wang2004image}
Z.~Wang, A.~C. Bovik, H.~R. Sheikh, and E.~P. Simoncelli, ``Image quality
  assessment: from error visibility to structural similarity,'' \emph{IEEE
  transactions on image processing}, vol.~13, no.~4, pp. 600--612, 2004.

\bibitem{liu2015deep}
Z.~Liu, P.~Luo, X.~Wang, and X.~Tang, ``Deep Learning Face Attributes in the Wild,'' in \emph{Proceedings of the IEEE International Conference on Computer Vision}, 2015, pp. 3730--3738.

\bibitem{everingham2015pascal}
M.~Everingham, S.~M.~A. Eslami, L.~Van Gool, C.~K.~I. Williams, J.~Winn, and A.~Zisserman, ``The PASCAL Visual Object Classes Challenge: A Retrospective,'' \emph{International Journal of Computer Vision}, vol.~111, no.~1, pp. 98--136, 2015.
\end{thebibliography}

\vfill

\end{document}